\documentclass{article}

\usepackage{PRIMEarxiv}

\usepackage[utf8]{inputenc} % allow utf-8 input
\usepackage[T1]{fontenc}    % use 8-bit T1 fonts
\usepackage{hyperref}       % hyperlinks
\usepackage{url}            % simple URL typesetting
\usepackage{booktabs}       % professional-quality tables
\usepackage{amsfonts}       % blackboard math symbols
\usepackage{nicefrac}       % compact symbols for 1/2, etc.
\usepackage{microtype}      % microtypography
\usepackage{lipsum}
\usepackage{fancyhdr}       % header
\usepackage{graphicx}       % graphics
\usepackage{caption}
\usepackage{subcaption}
\usepackage[sort&compress,round,comma,authoryear]{natbib}
\usepackage{amsmath,amssymb,array}
\usepackage{xspace}
\graphicspath{{images/}}     % organize your images and other figures under images/ folder

\title{\Twitmo: A Twitter Data Topic Modeling and Visualization Package for R
}

\author{
  Andreas Buchmüller, Gillian Kant, Christoph Weisser \\
  Centre for Statistics \\
  Humboldtallee 3, 37073 Göttingen\\
  University of Göttingen\\
  Göttingen, Germany\\
  \texttt{\{a.buchmueller, gillian.kant, c.weisser\}@stud.uni-goettingen.de} \\
   \And
  Krisztina Kis-Katos, Thomas Kneib \\
  Centre for Statistics \\
  Humboldtallee 3, 37073 Göttingen \\
  University of Göttingen\\
  Göttingen, Germany\\\
  \texttt{\{krisztina.kis-katos, tkneib\}@uni-goettingen.de} \\
     \And
  Benjamin Säfken\\
  Institute of Mathematics\\
  Erzstraße 1, 38678 Clausthal\\
  Clausthal University of Technology\\
  Clausthal, Germany\\
    
  \texttt{benjamin.saefken@tu-clausthal.de} \\
}

\begin{document}
\maketitle

\begin{abstract}
We present \texttt{Twitmo}, a package that provides a broad range of methods to collect, pre-process, analyze and visualize geo-tagged Twitter data. \texttt{Twitmo} enables the user to collect geo-tagged Tweets from Twitter and and provides a comprehensive and user-friendly toolbox to generate topic distributions from Latent Dirichlet Allocations (LDA), correlated topic models (CTM) and structural topic models (STM). Functions are included for pre-processing of text, model building and prediction. In addition, one of the innovations of the package is the automatic pooling of Tweets into longer pseudo-documents using hashtags and cosine similarities for better topic coherence \citep{hashtag-pooling}. 
The package additionally comes with functionality to visualize collected data sets and fitted models in static as well as interactive ways and offers built-in support for model visualizations via \texttt{LDAvis} providing great convenience for researchers in this area. The \texttt{Twitmo} package is an innovative toolbox that can be used to analyze public discourse of various topics, political parties or persons of interest in space and time.
\end{abstract}

% keywords can be removed
\keywords{NLP \and R \and Topic Modeling \and Twitter \and LDA \and STM \and CTM}

\section{Introduction}

% Intro & Outline
Twitter provides a rich source of publicly accessible text data in real time and a large variety of natural language processing (NLP) applications have been utilizing this data \citep{anber2016literature}. However, working with APIs can pose a technical challenge to many researchers, while topic modeling with the inherently noisy and sparse Tweets is a difficulty on its own. Still, with its 336 million daily active users creating 500 million Tweets per day \citep{ahlgren202040+}, the micro-blogging service Twitter offers NLP researchers access to a vast amount of geo-tagged text data. \texttt{Twitmo} provides all functions necessary for an NLP workflow including the sampling, pre-processing, analysis and visualisation of text data with a clear focus on usability.
The goal of this paper is to provide a summary overview of a topic modeling workflow with Twitter data in R. Additionally, we aim to highlight the main functions of the \texttt{Twitmo} package and demonstrate via an example how the package reduces the effort needed to produce coherent topic models with Twitter data, while also offering remedies for arising problems when modeling sparse and noisy text such as Tweets. We show that \texttt{Twitmo} handles such workflows much more efficiently than currently existing packages.

%% Twitter API und jsonlite/rjson und httr
Twitter provides access to its data through an API in two ways, namely by \emph{searching} or \emph{streaming}. Searching refers to the standard search access endpoint \citep{search-api-ref}, where researchers can collect and sample Tweets from the past nine days using keyword queries, latitude and longitude coordinates and radius or a combination of both.
The streaming API \citep{stream-api-ref} allows researchers to collect Tweets in real-time using keywords, additional filters like language, bounding box coordinates indicating a geographic location, a vector of Twitter user IDs, or simply to randomly sample a small amount of all publicly available data without filtering until the rate limit applies.
Since Twitters API follows REST principles, the endpoints will return data in the JSON format. Researchers need a JSON parser like \texttt{jsonlite} \citep{jsonlite-pkg} or \texttt{rjson} \citep{rjson-pkg} as well as a package to send and receive HTTP requests like \texttt{httr} \citep{httr-pkg}. Apart from text (sampled Tweets), Twitter returns a plethora of useful metadata, which can be used to enhance the analysis.
Researchers who want to sample Tweets one way or another must authenticate themselves by providing a so called bearer token. This requires the researchers first, to apply for a Twitter developer account, and second, to create a token through the developer portal. An application may be rejected if e.g., the explanation of the research project in question is not sufficient. This can be problematic since the exact question of a research project may not be completely clear before data is acquired since researchers need to familiarize themselves with the data first. Additionally, the application may take several weeks to be approved. 
In summary, working with and getting access to the API can be daunting at the beginning. \texttt{Twitmo} aims to be a remedy for this by leveraging the functionality of existing packages in the NLP space while also offering great usability and convenience.

%%% Comparison with existing tools
%% Packages für Twitter API
There exist multiple packages that aim to provide access to the Twitter API via R: these include \texttt{twitteR} \citep{twitteR-pkg}, \texttt{streamR} \citep{streamR-pkg}, and \texttt{rtweet} \citep{rtweet-package}.
%
%% Kurzer Absatz über die Vorteile und Limitierungen von rtweet
Since it is the only package actively maintained as of 2021, \texttt{rtweet} is the best way to start collecting Twitter data with R. \texttt{rtweet} eases the process of sampling by providing direct access to streaming and searching as well as numerous additional endpoints and comes with some built-in parsing capabilities. On top, \texttt{rtweet} does not require a developer account as it works with a regular Twitter account. Despite somewhat lower rate limits for certain endpoints (see \href{https://developer.twitter.com/en/docs/twitter-api/v1/rate-limits}{rate limits}), \texttt{rtweet} is a a great tool for researchers to familiarize themselves with Twitter data. Once a Developer Account is acquired, a bearer token can be supplied later on to increase rate limits, enhancing its capabilities even further by allowing to sample more data in less time. However, \texttt{rtweet} offers no way of analyzing and has visualization capabilities limited to time series plots of the sampled data.
\texttt{Twitmo} uses \texttt{rtweet} for collecting Tweets with a couple of added features to improve usability. For instance, \texttt{Twitmo} provides bounding boxes for 246 countries and regions as well as 249 cities, converts latitude and longitude coordinates to bounding box coordinates and allows for a convenient sampling of geo-tagged Tweets without having to rely on external tools.
%

%% Tweet Pooling und die Rolle vom Quanteda
Besides of sampling, parsing and pre-processing the data, corpus creation is another important part of the NLP workflow. There is an abundance of packages suitable for the analysis of text data, e.g., \texttt{tidytext} \citep{tidytext-pkg}, \texttt{tm} \citep{tm-paperref} as well as several different tokenizer packages like \texttt{tokenizers} \citep{tokenizers-pkg}, \texttt{udpipe} \citep{udpipe-pkg} or \texttt{spacyr} \citep{spacyr-pkg}. While all of these can handle text data in NLP analysis, these packages differ in the scope of their capabilities. Packages like \texttt{tokenizers} focus on doing one specific task---in this case the tokenization step, while packages like \texttt{udpipe} and \texttt{spacyr} provide fully annotated language models and state-of-the-art entity recognition. For researchers mainly interested in pre-processing, i.e., text cleaning, corpus creation, and tokenization of text, these packages provide too much overhead as pre-trained language models tend to be large and annotation and entity recognition is computationally expensive. \texttt{spacyr} additionally requires a working Python environment since it is an R wrapper arround the python package, not a native R package itself. 
A lean and flexible alternative is \texttt{quanteda} \citep{quanteda-pkg}. It offers a broad range of functions for text analysis including corpus creation and tokenization and comes with a large built-in set of methods for text cleaning, which can also be combined with external tokenizers like the \texttt{tokenizers} package. Written in R and C++, \texttt{quanteda} is very efficient, fast and actively maintained by its contributors. It also comes with built-in compatibility to packages like \texttt{tm}, \texttt{stm} \citep{stm-pkg}, \texttt{lda} \citep{lda-pkg} and \texttt{topicmodels}, \citep{topicmodels-pkg} making it extremely well suited for pre-processing tasks. Our package leverages the speed of \texttt{quanteda} by using it for corpus construction, some pre-processing tasks (like stopword removal, creation of n-grams, text cleaning) and calculating document frequency matrices (dfm) based on hashtags. By using dfm objects for our hashtag pooling, we ensure compatibility to a wide range of other R packages, while also efficiently handling the workload.

%% Übersicht Topic Modeling Packages
There are many packages suited for topic modeling tasks in R. For LDA models in particular there exist multiple packages like \texttt{topicmodels} \citep{topicmodels-pkg} and \texttt{lda} \citep{lda-pkg}, both of which have their underlying calculations performed in C, making them very efficient. Apart from LDA topic models, which are the most popular choice in topic modeling, there is also the native R package \texttt{stm} \citep{stm-pkg}, which can be used to build structural topic models (STM), an approach that incorporates text-level metadata. The \texttt{stm} package is available only in R and can be used especially for Twitter topic modeling since the Twitter API allows to access a large amount of metadata accompanying the Tweets themselves, e.g. the  date, location, favourite and like counts and many more.

While all of the above packages offer great functionality, they have drawbacks in terms of usability, especially when working with geo-tagged data like Tweets. \texttt{Twitmo} aims to leverage the functionality of currently existing packages in the NLP space by adding new functionality like hashtag-pooling and to improve their usability for Twitter topic modeling tasks by providing low-code options for visualizations and building coherent topic models.

%%% Kurzbeschreibung vom Package & potentielle use-cases
The \texttt{Twitmo} package provides a wide range of functions to collect, pre-process, analyze and visualize the contents of Twitter data.
The package allows the user to collect Tweets using a regular Twitter account for any area in the world by streaming and searching. Subsequently, the inherently noisy Twitter data can be cleaned, transformed to tabular data, saved and exported. For pre-processing \texttt{Twitmo} supports stopword removal in different languages (including Twitter specific stopwords like "amp" or "rt"), symbol removal, removal of emojis, removal of punctuation, removal of URLs as well as removal of hashtags and usernames from the corpus and the creation of n-grams.
Most importantly, \texttt{Twitmo} enables the user to build coherent topic models with extremely sparse Twitter data by preparing the Tweets for analysis. We pool Tweets using hashtags and cosine similarities as proposed by \cite{hashtag-pooling}, a technique that was shown to improve topical coherence of short-text LDA models and for which there is no existing R implementation yet. Our package also helps researchers make use of the metadata accompanying Tweets to build STMs.
\texttt{Twitmo} also provides options for automatized Topic Model parameter optimization, assisting researchers in finding the optimal hyperparameters for their models.
The distribution of topics over documents can be visualized with the popular \texttt{LDAvis} package, for which \texttt{Twitmo} provides full built-in compatibility. \texttt{Twitmo} can also be used to conveniently visualize sampled Tweets in an interactive \texttt{leaflet} \citep{leaflet-pkg} map without pre-processing.
Above this, the spatial distribution of Tweets can be plotted on a static world map, with support for all major regions of the world, in order to visualize the distribution of sampled Tweets. Spatial distribution plots can also be restricted to certain hashtags. In summary, \texttt{Twitmo} leverages the advantages of multiple R packages to enable researchers to build coherent topic models with geo-tagged Twitter data in R. It supports all parts of the workflow from sampling Tweets to generating topic distributions from Latent Dirichlet Allocation (LDA) topic models and Structural topic models (STM). The package can be used for a wide range of applications for scientific research in order to gain insights into topics discussed on Twitter and as such is an innovative toolbox tailored for the analysis of public discourse. For example \texttt{Twitmo} can be used to analyze the discourse in the weeks leading up to the election or on election night, modeling Covid-19 discussion, the spread of misinformation in certain areas of the world as well as monitoring and analyzing the discussions in disaster struck regions.

\section{Latent dirichlet allocation and structural topic models}
%% Theoretischer Hintergrund zu LDA und STM Modellen
This section describes the different topic modelling approaches available in the package as well as the pseudo-document generation process used.
An overview of the relevant quantities, notations and assumptions can be found in Table~\ref{variable_list}.

\begin{table}
\small
\begin{tabular*}{\linewidth}{@{\extracolsep{\fill}}*8l@{}} \toprule
  
$K$                   &   Number of topics                    \\

$D$                   &   Number of documents in a corpus                \\

\textit{d}                   &   Document            \\

$\boldsymbol{\beta}_{k}  \sim \mathrm{Dir}(\boldsymbol{\beta})$   &   Word distribution for topic $k$                \\

$\boldsymbol{\theta}_d \sim \mathrm{Dir}(\boldsymbol{\alpha})$    &   Topic distribution for document $d$             \\

$\boldsymbol{\beta}$  & Parameter of the Dirichlet prior on the per-topic word distribution \\

$\boldsymbol{\alpha}$  & Parameter of the Dirichlet prior on the per-document topic distributions \\

$N_{d}$ & Number of words for document $d$ \\

$z_{nd} \sim \mathrm{Multinomial}(\boldsymbol{\theta}_d)$  & Topic of the ${n}^{th}$ word in document $d$ \\

$w_{nd}\sim \mathrm{Multinomial}(\boldsymbol{\beta}_{z_{nd}})$ & $n^{th}$ word in document $d$  \\

\bottomrule
    \end{tabular*}
    \caption{Variable description for LDA}
    \label{variable_list}
\end{table}

\subsection{LDA topic models}
Latent Dirichlet Allocation (LDA) topic models are best known and widely used for detecting latent topics in text documents. Originally proposed by \cite{lda}, the LDA is a generative probabilistic model. We define the corpus as a collection of $D$ documents where each document is generated as a mixture of underlying topics, where the continuous-valued mixture proportions  are  distributed  as  a  latent  Dirichlet random variable. A  topic  is  then defined by a distribution over all words in the corpus. More precisely, LDA assumes that each document $d$ in a corpus consisting of $d=1,\ldots,D$ documents is generated as follows:

\begin{enumerate}
    \item Determine $K$ topic distributions as $\boldsymbol{\beta}_{k}  \sim \mathrm{Dir}(\boldsymbol{\beta})$ where ${\boldsymbol{\beta}}=({\beta}_1,\ldots, {\beta}_K)$ represents the word relevances in a topic $k$.
    %    \item Determine the number of words as $N_d\sim \mathrm{Poisson}(\eta)$ where the rate parameter $\eta>0$ represents the expected length of a document.
    \item Determine the distribution over topics for document $d$ as $\boldsymbol{\theta}_d \sim \mathrm{Dir}(\boldsymbol{\alpha})$ where ${\boldsymbol{\alpha}}=({\alpha}_1,\ldots, {\alpha}_K)$ represents the vector of topic relevances for the corpus.
    \item To generate the $N_d$ words $w_{nd}$, $n=1,\ldots,N_d$ for document $d$:
    \begin{enumerate}
        \item choose a topic $z_{nd} \sim \mathrm{Multinomial}(\boldsymbol{\theta}_d)$ and
        \item determine the corresponding words $w_{nd}\sim \mathrm{Multinomial}(\boldsymbol{\beta}_{z_{nd}})$ where $\boldsymbol{\beta}_{z}$ is the vector of word occurrence probabilities $p(w|z)$ given topic $z$.
    \end{enumerate}
\end{enumerate}

The hyperparameters of the LDA are the Dirichlet parameters $\boldsymbol{\beta}$ and $\boldsymbol{\alpha}$. Note that $\boldsymbol{\beta}$ consists of all topic-wise word occurrence probabilities, $\beta_{kn}$, while $\boldsymbol{\theta_d}$ contains all document-wise topic occurrence probabilities, $\theta_{dk}$, that can be interpreted as the probabilities that a document $d$ was generated by a topic $k$. Marginalizing over the latent topics, the generative process for the words of a document $d$ can be written as:

\begin{equation} 
p(w_{nd}|\boldsymbol{\theta}_d,\boldsymbol{\beta})  = \sum_{k=1}^{K}  p(w_{nd}|z=k,\boldsymbol{\beta})p(z=k|\boldsymbol{\theta}_d),
\end{equation}

indicating that the LDA model is a mixture model, where the word-specific multinomial models $p(w_{nd}|z,\boldsymbol{\beta})$ are the mixture components and the topic probabilities $p(z|\boldsymbol{\theta}_d)$  are the respective mixture weights. 

The generating process for a document $d$ can be written as a product of word probabilities $p(w_{nd}|\boldsymbol{\theta}_d,\boldsymbol{\beta})$ and integration over $\boldsymbol{\theta}_d$:

\begin{equation} 
p(d | \boldsymbol{\alpha},\boldsymbol{\beta})  = \int p(\boldsymbol{\theta}_d | \boldsymbol{\alpha}) (\prod_{n=1}^{N_d} \sum_{k=1}^{K} p(w_{nd}|z=k,\boldsymbol{\beta})p(z=k|\boldsymbol{\theta}_d)) d\boldsymbol{\theta}_d.
\end{equation}

The posterior distribution of the hidden variables can be estimated with Gibbs sampling or variational mean field Bayes \citep{lda}.

\subsection{Structural topic models}
Structural topic models (STM) have been proposed by \cite{stm} and can be considered to be an improvement and extension over LDA models. The generative process is very similar to LDA models, where a document is a mixture over topics and a document is a mixture over words but additionally STM incorporate some key advantages over LDA models, which make them particularly useful for Twitter topic modeling tasks. By overcoming limitations of LDA such as accounting for topic correlation (thus dealing with the assumption of topic independence in LDA) and adding document-level metadata (categorical or numerical) as additional information, STMs offer a remedy to some of the problems of previous approaches when modeling short texts. As such they are a valuable tool for modeling short texts, which are often found in social media.

\begin{table}
\small
%% Muss noch auf das notwendigste zusammengestrichen werden
\begin{tabular*}{\linewidth}{@{\extracolsep{\fill}}*8l@{}} \toprule
$V$ & Vocabulary of unique terms \\
$m_{v}$ & Marginal log frequency of term $v$ (baseline word distribution) \\
%$\mathbf{X_{D \times P}}$ & Matrix of topic prevalence covariates\\
$X_{d}$ & 1-by- $p$ vector of topic prevalances \\
%$\mathbf{Y_{D \times A}}$ & 	Matrix of topical content covariates \\
$y_d$ & 1-by- $a$ vector of ocument-level content covariate\\
%$s, r, \rho, \sigma$ & 	Scalar hyperparameters \\
$\boldsymbol\Gamma_{P \times (K-1)}=\left[\gamma_{1}|\ldots| \gamma_{K}\right]$ & Matrix of coefficients for the topic prevalence model\\
$\left\{\kappa_{\cdots}^{(t)}, \kappa_{\cdots}^{(c)}, \kappa_{\ldots i}^{(i)}\right\}$ & Collection of coefficients for topical content model \\
$\boldsymbol{\gamma}_{k} \sim \mathrm{Normal}_{P}\left(0, \sigma_{k}^{2} I_{P}\right) $ & Topical coefficients for topics\\
$\boldsymbol\beta_{dk} \propto \exp \left(m_v+\kappa_{k}^{(\mathrm{t})}+\kappa_{y_{d}}^{(\mathrm{c})}+\kappa_{y_{d}, k}^{(\mathrm{i})}\right)$ & Word distrubition for topic $k$ in document $d$\\
$\boldsymbol{\theta}_{d} \sim \mathrm{LogisticNormal}_{K-1}\left(\boldsymbol{\mu_{d}} = \boldsymbol{\Gamma}^{\prime} \mathbf{x}_{d}^{\prime}, \mathbf{\Sigma}\right)$ & Core language model \\
$\mathbf{z}_{nd} \sim \mathrm{Multinomial}_{K}\left(\boldsymbol{\theta}_{d}\right)$ & Topic of the ${n}^{th}$ word in document $d$ \\
$w_{nd}\sim \mathrm{Multinomial}(\boldsymbol{\beta}_{z_{nd}})$ & $n^{th}$ word in document $d$ \\
%$\boldsymbol\beta_{dkv}= \frac{\exp \left(m_{0}+\kappa_{k, v}^{(\mathrm{t})}+\kappa_{y_{j}, v}^{(\mathrm{c})}+\kappa_{y_{d}, k, v}^{(\mathrm{i})}\right)}{\sum_{v} \exp \left(m_{v}+\kappa_{k, v}^{(\mathrm{t})}+\kappa_{y_{d}, v}^{(\mathrm{c})}+\kappa_{y_{d}, k, v}^{(\mathrm{i})}\right)}$ &  Full topical content model\\
$\eta_{d} \sim$ $\mathrm{Normal}_{K-1}\left(\boldsymbol{\mu}_{d}, \mathbf{\Sigma}\right)$ & Gaussian RV set to 0 for identification purposes\\
$\boldsymbol\theta_{dk}=\exp \left(\eta_{dk}\right) /\left(\sum_{i=1}^{K} \exp \left(\eta_{di}\right)\right)$ &	 Logistic-normal topic proportions of document $d$ for topic $k$\\

\bottomrule
    \end{tabular*}
    \caption{Variable description for STM}
    \label{variable_list_stm}
\end{table}

Recall from the LDA model that each document $d$ in a corpus consists of  $k=1,\ldots,K$ topics and words within the documents that are indexed by $n=1, \dots, N_d$. Additionally, observations consist of words $w_{nd}$ that are occurrences of unique terms from a document-level vocabulary of terms $v =1, \dots, V$. Assuming there are $k$ latent topics within the document corpus, the data generating process for each document $d$, given the number of topics $K$, with observed vocabulary of words $V$, observed words $w_{nd}$, the vector of document metadata $X_d$ (or design matrix $\mathbf{X_{D \times P}}$) and topical content vector $y_d$ (or design matrix $\mathbf{Y_{D \times A}}$) %as well as scalar hyperparameters $s, r, \rho$, and $K$ -dimensional hyperparameter vector $\sigma$, 
can be summarized as follows:

\begin{enumerate}
    \item Determine topic distributions from a logistic-normal generalized linear model $${\boldsymbol\theta}_{d} \mid X_{d} \gamma, \Sigma \sim \mathrm{LogisticNormal}\left(\mu_d=X_{d} \gamma, \Sigma\right)$$ where $\gamma$ is a $p$ -by- $K-1$ matrix of coefficients and $\Sigma$ is $K-1$ -by +$K-1$ covariance matrix.
    
    \item Determine the document-specific distribution over words for each topic $k$, as $$\beta_{dk} \propto \exp \left(m_v+\kappa_{k}^{(\mathrm{t})}+\kappa_{y_{d}}^{(\mathrm{c})}+\kappa_{y_{d}, k}^{(\mathrm{i})}\right)$$using the marginal log-frequency $m_v$ of term $v$, the topic specific deviation $\kappa_{k}^{(\mathrm{t})}$, the covariate group deviation $\kappa_{y_{d}}^{(\mathrm{c})}$ and the interaction between the two $\kappa_{y_{d}, k}^{(\mathrm{i})}$, where $m_v$ and $\kappa_{k}^{(\mathrm{t})}, \kappa_{y_{d}}^{(\mathrm{c})}$ and $\kappa_{y_{d}k}^{(\mathrm{i})}$ are 1-by-$V$ dimensional vectors with one item per unique term in the vocabulary. When no content covariates are present, $\boldsymbol\beta$ can be expressed as $\boldsymbol\beta_{dk} \propto$ $\exp \left(m+\kappa_{k}^{(\mathrm{t})}\right)$, in which case the model reduces to a correlated topic model (CTM) \citep{blei2007correlated}.

    \item Generate $N_d$ words $w_{nd}, n = 1, \dots, N_d$  for each document $d$:

\begin{enumerate}
    \item choose a topic based on the document-specific distribution over topics $z_{dn} \mid {\boldsymbol\theta}_{d} \sim \mathrm{Multinomial}\left({\boldsymbol\theta}_{d}\right)$

    \item determine the words $w_{nd} \mid z_{nd}, \boldsymbol\beta_{dk=z_{nd}} \sim \mathrm{Multinomial}\left(\boldsymbol\beta_{dk=z_{nd}}\right)$ conditional on the chosen topic
\end{enumerate}
\end{enumerate}

The generative process starts with document-topic and topic-word distributions generating documents but unlike in the LDA framework, documents have additional metadata $X_d$ associated with them, referred to as topical prevalence covariates. Given the topic mixtures $\boldsymbol\theta_{d}$, for each word within document $d$, a topic is sampled from a multinomial distribution $\mathbf{z}_{nd} \sim \operatorname{Multinomial}\left(\boldsymbol{\theta}_{d}\right)$. Conditional on the sampled topic, a word is drawn from the appropriate distribution over terms $\boldsymbol{\beta}_{z_{nd}}$. Contrary to LDA where both $\boldsymbol\alpha$ and $\boldsymbol{\beta}_{z_{nd}}$ are global parameters shared by all documents, in the STM framework $\boldsymbol\mu_d$ and $\boldsymbol{\beta}_{z_{nd}}$ are instead specified as functions of document-level covariates, which allow the vocabulary to vary for each document.

The  posterior  distribution  of  the  hidden  variables cannot be estimated via Gibbs sampling or  mean field variational Bayes due to nonconjugacy of the logistic-normal distribution used. Instead, a variant of nonconjugate variational expectation-maximization (VEM) can be used \citep{stm-long}. The full posterior of hidden variables can be expressed as:

$$
\begin{aligned}
p(\eta, \mathbf{z}, \kappa, \gamma, \Sigma \mid \mathbf{w}, \mathbf{X}, \mathbf{Y}) \propto 
&\left(\prod _ { d = 1 } ^ { D } \mathrm { N o r m a l } ( \eta _ { d } | { X } _ { d } \gamma , \Sigma ) \left(\prod_{n=1}^{N} \mathrm{Multinomial}\left(z_{nd} \mid \boldsymbol{\theta}_{d}\right)\right.\right. \\
&\left.\left.\quad \times \mathrm{Multinomial}\left(w_{n} \mid \boldsymbol{\beta}_{dk=z_{nd}}\right)\right)\right) \times \prod p(\kappa) \prod p(\boldsymbol{\Gamma})
\end{aligned}
$$
where $\eta_{d} \sim$ $\mathrm{Normal}_{K-1}\left(\boldsymbol{\mu}_{d}, \mathbf{\Sigma}\right)$ is a gaussian random-variable set to 0 for identification purposes and linked to the logistic-normal of $\boldsymbol\theta_d$ by specifying $\boldsymbol\theta_{dk}=\exp \left(\eta_{dk}\right) \mathbin{/} \left(\sum_{i=1}^{K} \exp \left(\eta_{di}\right)\right)$.

\subsection{Pseudo-document generation}
Recent research has shown that LDA may not perform well when handling short and sparse text data \cite{Maz16}, such as Tweets. Because Tweets are often not a mixture of topics but consist of one topic only, they are not fulfilling the LDAs main assumption, see \cite{Alv16}. We offer a remedy for this problem in the form of pooling in order to create longer pseudo-documents \cite{hashtag-pooling}. Pooling can be done by a feature that all documents share. By applying pooling, Tweets can be used to build coherent LDA topic models. For Tweets, \cite{kant2020ttlocvis} provides a pooling implementation that uses the hashtags of Tweets.
In \texttt{Twitmo}, we implemented hashtag pooling as proposed by \cite{hashtag-pooling}. Additionally, Tweets without hashtags can also be pooled into pseudo-documents by using cosine similarities calculated by their TF-IDF scores. The process can be summarized as follows:

\begin{enumerate}
    \item Determine unique hashtags
    \item For each unique hashtag append Tweets into longer documents. Tweets with multiple hashtags join multiple pools.
    \item For each created pseudo-document calculate a TF-IDF matrix and
    \begin{enumerate}
        \item for each unpooled Tweet calculate a TF-IDF matrix and
        \item calculate the pairwise cosine similarities between each and document pool and unpooled Tweet
        \item Append any unpooled Tweets, which passes a preset cosine threshold to the corresponding document pool
    \end{enumerate}
\end{enumerate}

The cosine threshold can be considered the only hyperparameter in this process. Since it is arbitrarily chosen, it might require some tuning by the researcher. Low cosine thresholds lead to longer pseudo-documents but less coherent topics, while simultaneously increasing computation time. When using too low thresholds, the package will return a warning. We recommend cosine thresholds between 0.5 and 0.9 for best results \citep{hashtag-pooling}. If the pools are too small, the cosine threshold should be gradually decreased until the pools can be used to model coherent topics. 

\section{Case Study} 
\subsection{Sampling geo-tagged tweets}

\texttt{Twitmo} can be used to collect Tweets with a regular Twitter account. As previously stated, \texttt{Twitmo} uses \texttt{rtweet} for streaming and searching Tweets, so data collected with \texttt{rtweet} can be analyzed and visualized with \texttt{Twitmo}.
Compared to \texttt{rtweet} our packages has a clear focus on usability and convenience when it comes to sampling geo-tagged Tweetsand offers additional features not included in \texttt{rtweet}, which focus on this purpose.
We will demonstrate this by streaming geo-tagged data from different regions of the world using \texttt{Twitmo}.

\begin{example}
    R> library(Twitmo)
    R> 
    R> get_tweets(method = 'stream',
    +            location = "DEU",
    +            timeout = 60,
    +            file_name = "tweets_from_germany.json")
\end{example}

In this example we use the streaming endpoint of the Twitter API to collect Tweets from Germany in real time for 60 seconds i.e. 1 minute. The Tweets are written to the working directory in the \code{"tweets\_from\_germany.json"} file. The location is specified with the location parameter \code{"DEU"}, a character string of a three letter country code following the ISO3166 alpha-3 standard. \texttt{Twitmo} has a database of bounding boxes of 246 locations and regions which can be accessed by typing \code{Twitmo:::bbox\_country}. This allows for very convenient sampling of geo-located data without relying on an external tool to draw a bounding box beforehand.
Locations also include cities. The following example live streams Tweets from New York City. A full list of supported cities can seen by using \code{rtweet:::citycoords}.

\begin{example}
    R> get_tweets(method = 'stream',
    +            location = "new york us",
    +            timeout = 60,
    +            file_name = "tweets_from_newyorkcity.json")
\end{example}

If more flexibility is needed, a custom bounding box can still be provided in the form of a vector of coordinates i.e. \code{c(southwest.longitude, southwest.latitude, northeast.longitude, northeast.latitude)}. For example a bounding box for the US mainland would look like:

\begin{example}
    R> get_tweets(method = 'stream',
    +            location =   c(-125, 26, -65, 49),
    +            file_name = "tweets_from_us_mainland.json")
\end{example}

If the location is not found within the package, nor a bounding box is specified the package will ask for a Google Maps API key to perform a lookup. If none is provided, the function will return a useful message pointing to the appropriate tables with valid locations is returned.

\subsection{Searching tweets}
Streaming in real-time is the best way to collect geo-tagged text. However, \texttt{Twitmo} also supports searching for Tweets. Searching is done for the last 9 days and can be done by specifying \code{method = 'search'}. Currently searching is only supported for cities and not countries. Searching also can be done by providing keywords in addition or without a location. Note that depending on the keyword and location, searching may not yield a significant enough amount of Tweets for a further analysis. But it can still be a valuable tool for trying to sample from past events. Here is a brief example of searching Tweets around the Berlin-Brandenburg metropolitan area limiting the search to 100 results.

\begin{example}
    R> get_tweets(method = 'search',
    +            location = "berlin",
    +            file_name = "tweets_from_berlin.json")
\end{example}

\subsection{Building a topic model with real data} 
%%% Diese Sektion ist reserviert für "Betriebsanweisungen"
% Ein volles Beispiel wird einmal durchexerciert um den Umfang und die Möglichkeiten des Package zu demonstrieren
We will demonstrate the capabilities of \texttt{Twitmo} by loading a sample piece of Twitter data containing $N=193$ Tweets from the US mainland, which is included in the package.

\begin{example}
    R> library(Twitmo)
    
    R> raw_path <- system.file("extdata", "tweets_20191027-141233.json", package = "Twitmo")
    
    R> mytweets <- load_tweets(raw_path)
    opening file input connection.
    Imported 193 records. Simplifying...
    closing file input connection.
\end{example}

The function takes a JSON file as input. It does not matter whether the JSON file has been sampled using \texttt{rtweet}, \texttt{Twitmo} or for instance via a python package like \emph{TTLocVis} \citep{kant2020ttlocvis}. 
Alternatively, the JSON file of interest can be loaded via \code{jsonlite::stream\_in()} in conjunction with \code{rtweet::tweets\_with\_users()}. The latter method only works, however, if the Tweets have been sampled with \texttt{rtweet} 0.7.0 or higher or \texttt{Twitmo}.
After loading the data with the \code{load\_tweets()} function, a data frame of shape $N \times 93$ is created, where $N$ is the number of Tweets in the file and $93$ is the number of columns, i.e. covariates after parsing the data.
Now the data frame can be used for hashtag-pooling with \texttt{Twitmo}.

\begin{example}
    R> pool <- pool_tweets(data = mytweets,
    +                     remove_numbers = TRUE,
    +                     remove_punct = TRUE,
    +                     remove_symbols = TRUE,
    +                     remove_url = TRUE,
    +                     remove_emojis = TRUE,
    +                     remove_users = TRUE,
    +                     remove_hashtags = TRUE,
    +                     cosine_threshold = 0.7,
    +                     stopwords = "en",
    +                     n_grams = 1)
    
    193 Tweets total
    158 Tweets without hashtag
    Pooling 35 Tweets with hashtags #
    56 Unique hashtags total
    Begin pooling ...Done
\end{example}

The \code{pool\_tweets()} function can be supplied with a wide range of pre-processing parameters as well as the pooling hyperparameter \code{cosine\_threshold}, which sets the similarity threshold at which unpooled Tweets will be appended to the corresponding document pools passing the user-specified threshold, as mentioned before. First, one can choose between different pre-processing parameters like removal of URLs, emojis, numbers, Twitter usernames and hashtags and more. One can also choose language specific stopwords, using English or any other language included in the \texttt{stopwords} package. Twitter-specific stopwords like "amp" will be appended to the stopword dictionary automatically, no matter which language is chosen. Additionally, n-grams can be specified via the \code{n\_grams} parameter. The cosine threshold parameter will determine the similarity of the Tweets without hashtags that will join the document pools.

\begin{example}
    R> summary(pool)
                         Length Class  Mode     
    meta                    12  tbl_df list     
    hashtags                 2  tbl_df list     
    tokens                  48  tokens list     
    corpus                  48  corpus character
    document_term_matrix 17232  dfm    S4   
\end{example}

Upon completion, \code{pool\_tweets()} will print a short description of the loaded JSON file as well as return a list with five objects: A tibble of metadata, a tibble of hashtags and their frequencies, a list of tokens in the corpus, the corpus itself where the hashtags are the document names as well as a dfm object, which can be used to build models later. The dfm object also holds metadata on a document-level, which makes it easy to build STM models later on.

LDA models can be fit via the \code{fit\_lda()} function. 

\begin{example}
    R> model <- fit_lda(pool$document_term_matrix, n_topics = 7, method = "Gibbs")
\end{example}

The \code{fit\_lda()} function takes three arguments of which the first two need to be specified. First, a dfm object from the pooled documents, which was returned by the \code{pool\_tweets} function earlier. The second argument, \code{n\_topics}, determines the number of topics $k$. The \code{"method"} parameter refers to the method used to obtain the posterior function, i.e. variational expectation maximization or Gibbs sampling.
Analogously, CTM models can be fitted via the \code{fit\_ctm()} function, which works the same as \code{fit\_lda()}.

\begin{example}
    R> model.ctm <- fit_ctm(pool$document_term_matrix, n_topics = 7)
\end{example}

Finding the optimal number of topics in a corpus is an iterative process. One way to determine the optimal number of topics is via the \code{find\_lda()} function. 

\begin{example}
    R> find_lda(pool$document_term_matrix, search_space = seq(1, 10, 1),  method = "Gibbs")
\end{example}

Using the \code{search\_space} parameter, multiple LDA models will be fitted and compared in terms of different metrics, which we will only briefly describe. 
\begin{enumerate}
    \item \cite{Arun2010} try to minimize KL-Divergence of distributions derived from matrix factors of the corpus. 
    \item \cite{CaoJuan2009} select the best LDA model by comparing topical density. 
    \item \cite{deveaud2014} use a simple heuristic that estimates the number of latent concepts of a user query by maximizing the information divergence between all pairs of LDAs topics in the search space. 
    \item \cite{Griffiths2004} use an MCMC algorithm for model selection and have shown that it works on scientific documents.
\end{enumerate}

From our experience with Twitter data, the approaches by \cite{Arun2010} and \cite{CaoJuan2009} tend to work better than the others. However, this differs depending on the corpus size and thus with regard to the length of the pseudo-documents created.
The function returns a plot, where the metrics are compared for each topic number $k$. This can be useful at the beginning of a new research project to gain insight into what $k$ might be. 
It is important to note that changing the \code{method} parameter also produces different results, so if a model is chosen based on its metrics when using Gibbs sampling, the \code{fit\_lda()} function needs to be supplied with the \code{method="Gibbs"} parameter, otherwise the fitted model might be different from the model fitted via \code{find\_lda()}.

After fitting an LDA model, \texttt{Twitmo} offers a variety of convenience functions, which make it easy to inspect the built models.
\code{lda\_terms()} provides the most likely $n$ terms (in terms of probability) for each of the $k$ topics of an LDA model. As such, \code{lda\_terms()} relates to the $\boldsymbol\beta$ parameter of our LDA model.

\begin{example}
    R> lda_terms(model, n_terms = 5)
      Topic.1 Topic.2  Topic.3   Topic.4    Topic.5       Topic.6 Topic.7
    1     bio     see      job       big      click          link    like
    2    last     job     life      love      puppy tenrestaurant     aka
    3   first     bio       us       fun      place         today   swipe
    4   photo patient   hiring gentleman       says       morning    meet
    5  bombed  sunday downtown       lol photoshoot        church     day
\end{example}

This can also be applied to hashtags with the \code{lda\_hashtags()} function, which will return a vector of the most likely (highest probability) topics for each hashtag.

\begin{example}
    R> lda_hashtags(model) %>% head(5)
                  Topic
    mood              2
    motivate          5
    healthcare        3
    mrrbnsnathome     6
    newyork           1
\end{example}  

Another way to explore the distribution of the fitted LDA is by using the \code{lda\_distribution()} function. The function will return either the term distribution over topics $\beta$ or the document distribution over topics $\gamma$. Gamma is an additional parameter that was not mentioned in the original LDA specification but is included in the \texttt{lda} package \citep{lda-pkg} and thus supported by \texttt{Twitmo}. It refers to the document distribution over topics, i.e. the probability of a hashtag belonging to a certain topic. If \code{tidy = TRUE}, a tibble will be returned instead of a data frame.

\begin{example}
    R> lda_distribution(model, param = "gamma", tidy = TRUE)
    # A tibble: 48 × 8
       document            `1`   `2`   `3`   `4`   `5`   `6`   `7`
       <chr>             <dbl> <dbl> <dbl> <dbl> <dbl> <dbl> <dbl>
     1 about             0.149 0.134 0.105 0.193 0.134 0.134 0.149
     2 beautifulskyz     0.169 0.154 0.199 0.108 0.108 0.123 0.139
     3 bonjour           0.133 0.133 0.150 0.117 0.150 0.150 0.166
     4 breakfast         0.163 0.128 0.128 0.128 0.128 0.181 0.145
     5 chinup            0.135 0.154 0.135 0.135 0.154 0.154 0.135
\end{example}

After inspecting the fitted models, the user may continue with predictions.
Predictions are implemented in the \code{predict\_lda()} function. The function may be applied to the original data frame of Tweets (before pooling) as well as on out-of-sample Tweets as long as they have been properly parsed with \texttt{Twitmo} or \texttt{rtweet}.

\begin{example}
    R> predict_lda(mytweets, model, response = "max") %>% tail(5)
    text189 text190 text191 text192 text193 
          7       2       2       5       1
          
    R> predict_lda(mytweets, model, response = "prob") %>% tail(5)
                    1         2         3         4         5         6         7
    text189 0.1662338 0.1298701 0.1662338 0.1298701 0.1480519 0.1298701 0.1298701
    text190 0.1513859 0.1812367 0.1812367 0.1066098 0.1215352 0.1215352 0.1364606
    text191 0.1098901 0.1560440 0.1560440 0.1252747 0.1252747 0.1714286 0.1560440
    text192 0.1235521 0.1235521 0.1370656 0.1370656 0.2181467 0.1640927 0.0965251
    text193 0.1133787 0.1292517 0.1927438 0.1133787 0.1451247 0.1451247 0.1609977
\end{example}

The latter option returns topic probabilities for each text (Tweet). Predictions can be given as point estimators (\code{response="max"}) or probabilities (\code{response="prob"}). 

\subsection{Black- and whitelisting keywords}
As mentioned earlier, one way to build coherent topics models is through hyperparameter tuning, namely by trying different $k$ with the \code{find\_lda()} function. Another way of arriving at a coherent topic model is through optimal pre-processing of the Tweets.
Certain keywords may heavily dilute or worsen topic coherence. As a remedy for this, \texttt{Twitmo} comes with a function for black- and whitelist certain keywords from the collected Tweets. The \code{filter\_tweets()} function takes a data frame of loaded Tweets and a keyword dictionary containing a character string of keywords to be used for black- or whitelisting. In the brief example below, we blacklist Tweets with mentions of "football" and "mood" in them. We then use the \code{dim()} function to verify that we have successfully reduced our data frame by 10 rows, i.e. excluded 10 Tweets from our set.

\begin{example}
    R> filter_tweets(mytweets, keywords = "football,mood", include = FALSE) %>% dim()
    [1] 183  93
    R> mytweets %>% dim()
    [1] 193  93
\end{example}

\subsection{Building a structural topic model}
When dealing with a small sample of Tweets, building a coherent LDA topic model might be difficult due to the short- and noisiness of the text. In this case building a structural topic model could provide better results. Since STMs leverage the metadata accompanying Tweets they are a good alternative to the classic LDA approach. 
Similarly to LDA topic models, STMs can be fit via the \code{fit\_stm()} function. Since STMs are capable of incorporating external covariates, e.g., favourite or retweet counts, the function comes with a \code{xcov} parameter, which takes a vector or matrix of size $N \times P$, where $P$ is the number of external covariates, as input. External covariates must be supplied for every Tweet, so the length of the metadata vector matches the number of Tweets in the data frame. 
We will demonstrate this by using the US mainland data set included in the package.

\begin{example}
    R> summary(pool)
                         Length Class  Mode     
    meta                    12  tbl_df list     
    hashtags                 2  tbl_df list     
    tokens                  48  tokens list     
    corpus                  48  corpus character
    document_term_matrix 17232  dfm    S4  
\end{example}

Pooling the data returns a named list with a data frame called "meta" inside. This is a pre-selection of metadata that works with the \texttt{stm} package used for the underlying calculations and we consider most important. Any covariate can be used as long as it is supported by the \texttt{stm} package.
In this example, we are fitting a STM with 7 topics and use the retweet count, friends count, followers count, reply count, quote count, favorite count and the date as external covariates for each Tweet. Before using providing the date we make sure it is a date time object and not a character string. Since STMs only work with unpooled (parsed) Tweets (because it makes little sense to pool metadata) pre-processing is done before fitting via their respective parameters inside the \code{fit\_stm} function.
The topic model can be then be inspected via \code{summary(stm.model)}. The \code{stm.model} object will contain a named list "prep" containing the pre-processed documents of the model, which can be reused to avoid pre-processing each time.

\begin{example}
    R> stm.model <- fit_stm(mytweets, n_topics = 7, 
    +                      xcov = ~ retweet_count + followers_count + reply_count + 
    +                      quote_count + favorite_count,
    +                      remove_punct = TRUE,
    +                      remove_url = TRUE,
    +                      remove_emojis = TRUE,
    +                      stem = TRUE,
    +                      stopwords = "en")
    Building corpus... 
    Converting to Lower Case... 
    Removing punctuation... 
    Removing stopwords... 
    Remove Custom Stopwords...
    Removing numbers... 
    Stemming... 
    Creating Output... 
    Removing 978 of 1302 terms (978 of 2112 tokens) due to frequency 
    Your corpus now has 137 documents, 324 terms and 1134 tokens. 
    Beginning Spectral Initialization 
    	 Calculating the gram matrix...
    	 Finding anchor words...
     	.......
    	 Recovering initialization...
     	...
    Initialization complete.
\end{example}

External covariates (metadata) can be provided via the \code{xcov} parameter. The \code{xcov} parameter takes a formula object with an empty left-hand side, a character string or a character vector as input and is more flexible than the \code{prevalence} parameter from the the \texttt{stm} packages \code{stm} function (to which covariate names are passed) in this regard. The user needs only to make sure that the chosen external covariates are present as columns in the data frame of parsed Tweets.

\begin{example}
    R> stm.covariates <- "retweet_count,followers_count,reply_count,quote_count,favorite_count"
    
    R> stm.model <- fit_stm(mytweets, n_topics = 7, xcov =  stm.covariates)
\end{example}

While they are a good alternative to LDA topic models for smaller data sets, STMs are computationally expensive. They are unsuitable for larger data sets, especially when working with less powerful machines, but offer a valuable tool for working with smaller data sets where it might be difficult to build coherent LDA topic models.

\subsection{Visualizing tweets and models}

\texttt{Twitmo} also comes with built-in static and interactive plotting methods.
Static plots can be created with the \code{plot\_tweets()} function and will return a static map with a data point for each tweet in the data set. The usage of the \code{alpha} parameter can be used to change to opacity of data points for optimal results. Larger data sets require lower alpha values.

To illustrate, we plot a static map of our included US data set as well as a data set containing 22,000 Tweets from the UK (which are not included in the package).
Static plotting can be applied to all regions of the world. For this, the \code{region} argument needs to be passed with a valid character vector encoded in ISO3166. A data frame of all valid character vectors is included in the \texttt{maps} package and can be accessed by using \code{maps::iso3166}. In our US mainland example we used the \code{region = "USA(?!:Alaska|:Hawaii)} argument, specifying the USA without Alaska and Hawaii, i.e., only the US mainland.

\begin{example}
    R> mytweets <- subset(mytweets, country_code == "US")
    
    R> plot_tweets(mytweets, region = "USA(?!:Alaska|:Hawaii)", alpha=0.1)
\end{example}

\begin{figure}[h]
    \centering
    \includegraphics[width=1\textwidth]{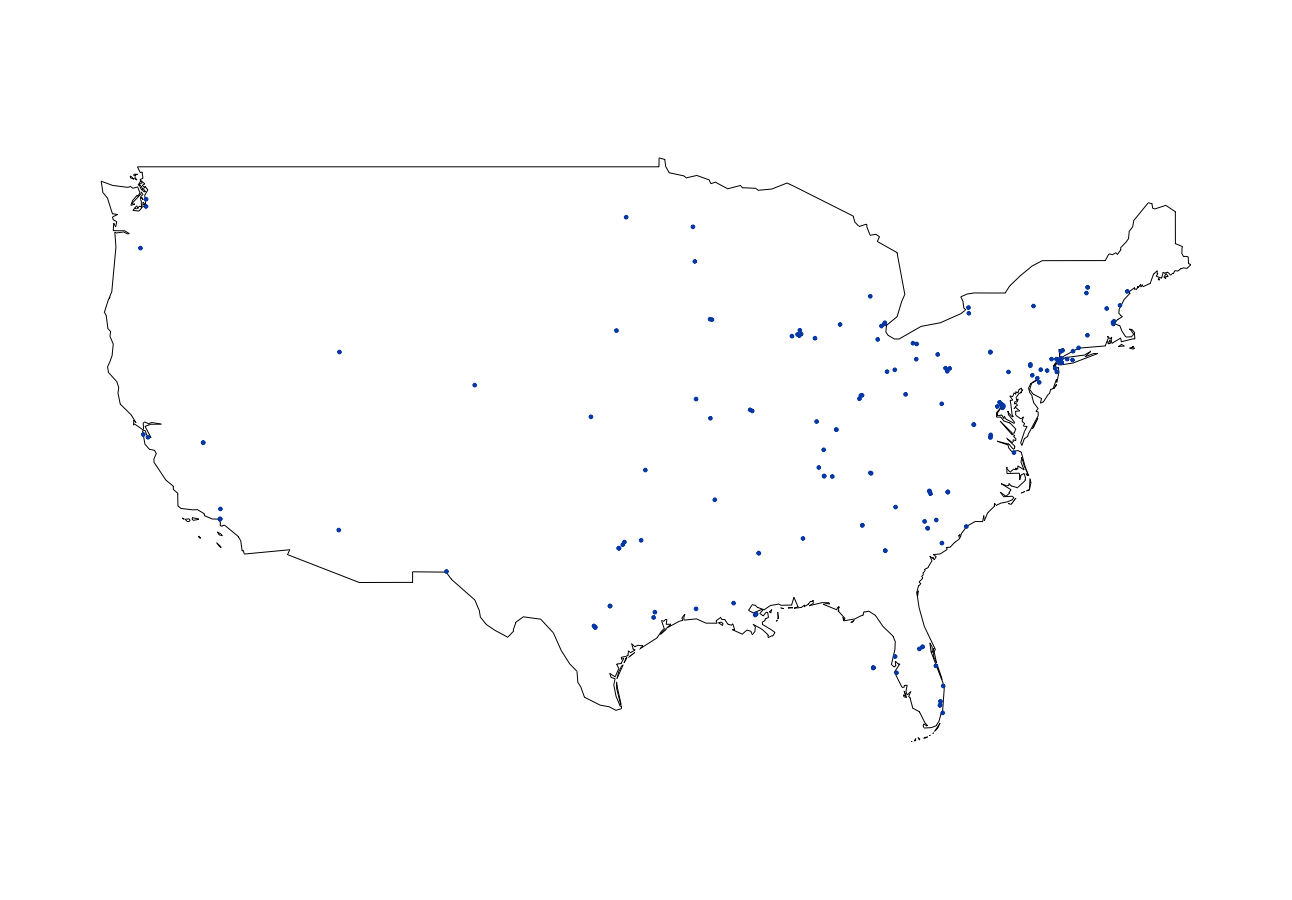}
    \caption{Distribution of tweets for US mainland data set}
    \label{fig:ttvis_static_us}
\end{figure}

Using geo-tagged Tweets from the UK, which are much denser compared to the US mainland data, illustrates that reducing the \code{alpha} value of \code{plot\_tweets()} increases visibility in larger samples. The region is set to the region specific ISO3166 alpha-3 country code for the United Kingdom \code{"UK"}. Before we plot our Tweets, we make sure that we do not plot Tweets from neighbouring states, e.g. Ireland, by subsetting the Tweets using the two-letter country code for the United Kingdom.

\begin{example}
    R> uk_tweets <- load_tweets("~/fromUK.json")
  
    R> uk_tweets <- subset(uk_tweets, country_code == "GB")
  
    R> plot_tweets(uk_tweets, region = "UK", alpha = 0.1)
\end{example}

\begin{figure}[h]
\centering
\begin{subfigure}{.5\textwidth}
    \centering
    \includegraphics[width=.5\linewidth]{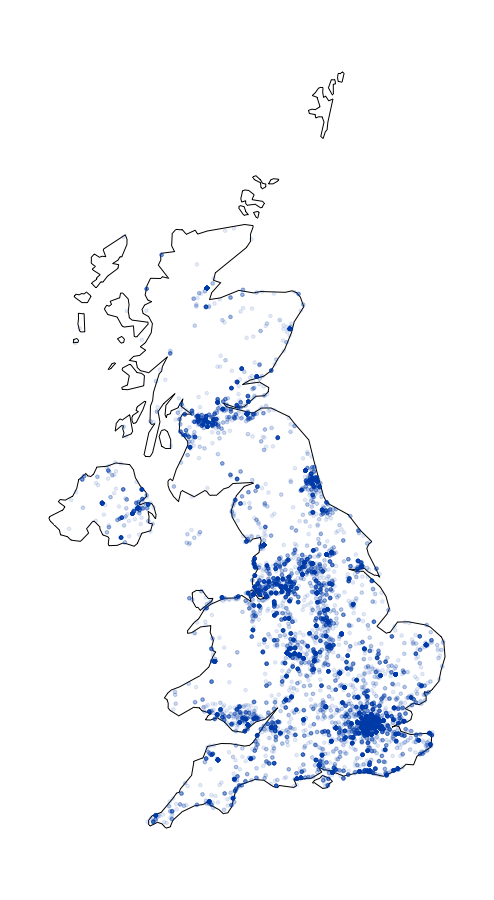}
    \caption{Distribution of tweets}
    \label{fig:map_uk_highres}
\end{subfigure}%
\begin{subfigure}{.5\textwidth}
    \centering
    \includegraphics[width=.5\linewidth]{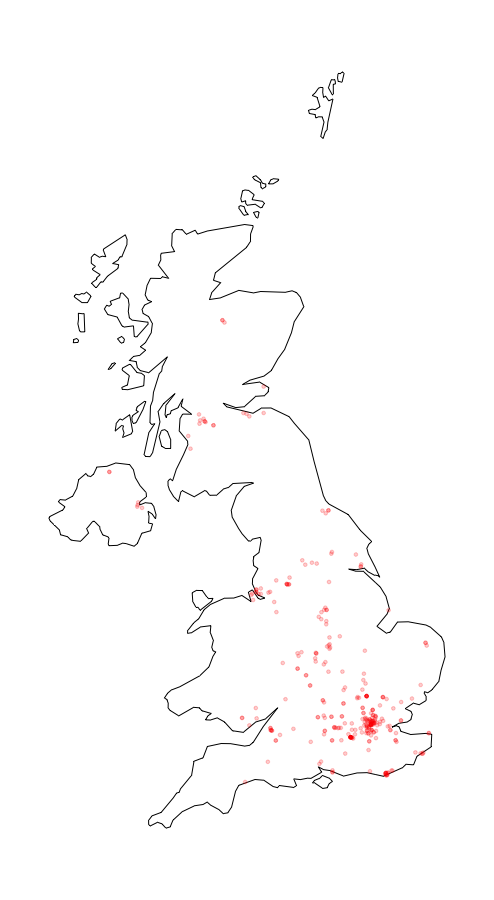}
    \caption{Distribution of the "foodwaste" hashtag}
    \label{fig:ht_map_uk}
\end{subfigure}
\caption{Tweet and hashtag plots for UK mainland data set}
\end{figure}

Static plotting also works with hashtags. The \code{plot\_hashtags()} function works in a similar fashion as the \code{plot\_tweets()} function but supports two additional arguments. The \code{hashtag} argument is the filter for the hashtag to be displayed. With the \code{case\_sensitive} argument the user can turn case sensitivity either on or off for the character string in \code{hashtag}. Although usually in pre-processing all words are converted to lower case, in certain occasions case sensitivity can carry a meaning---for instance, in certain popular memes on Twitter---and thus might be a valuable source of information. For illustration, we begin by inspecting the hashtag counts in the UK mainland data and then plot the distribution of the third most common hashtag "foodwaste" on the map using an \code{alpha} value of 0.2 shown in Figure \ref{fig:ht_map_uk}.

\begin{example}  
 
    # plot the "foodwaste" hashtag
    R> pool_uk <- pool_tweets(data =  uk_tweets)  
    R> pool_uk$hashtags %>% head(5)
    
    # A tibble: 5 x 2
    hashtag       count
    <chr>         <int>
    1 unitedkingdom   367
    2 free            366
    3 foodwaste       321
    4 london           91
    5 thechase         72
    
    R> plot_hashtag(uk_tweets, region="UK", hashtag="foodwaste", ignore_case=TRUE, alpha=0.2)
\end{example}

\begin{figure}[h]
    \centering
    \includegraphics[width=1\textwidth]{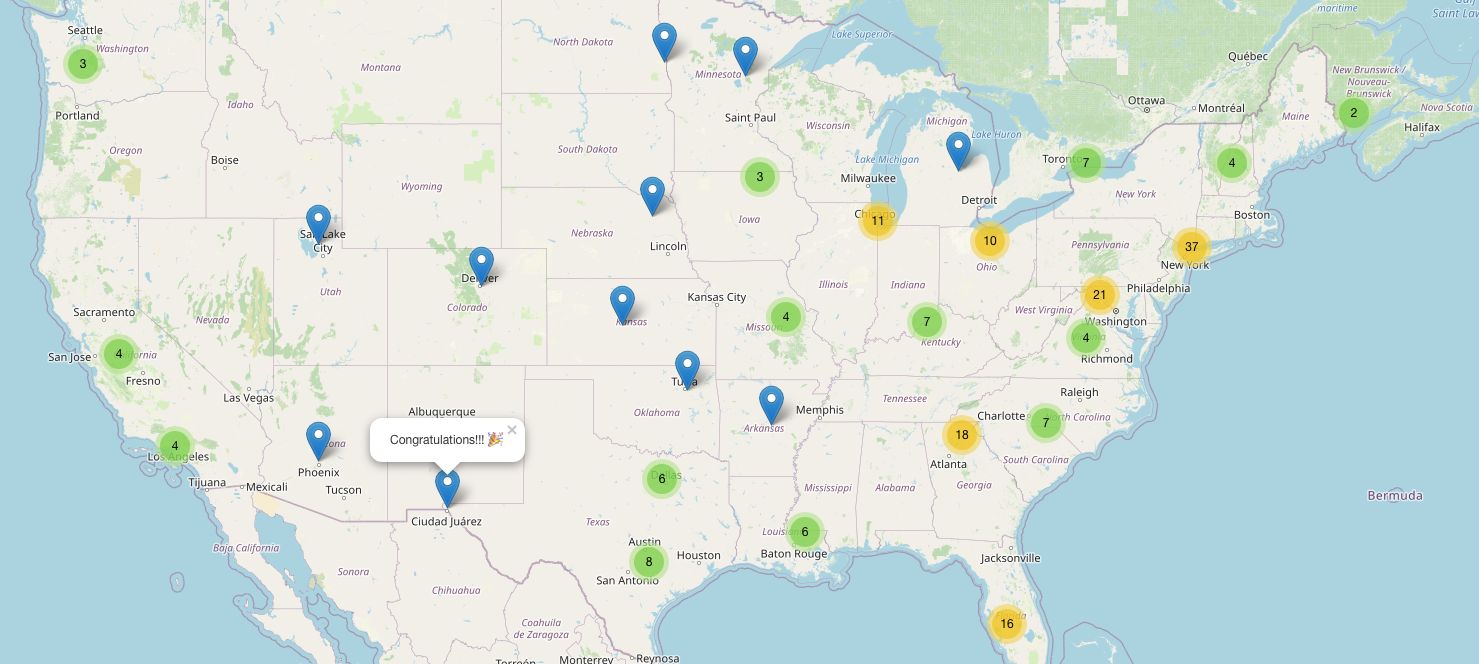}
    \caption{Interactive Map of US mainland Tweets}
    \label{fig:ttvis_interactive_us_crop}
\end{figure}

Another way of visualizing geo-tagged Twitter data is through the \code{cluster\_tweets()} function. It creates an interactive \texttt{leaflet} map of loaded geo-tagged Tweets with only one line of code depicted in Figure \ref{fig:ttvis_interactive_us_crop}. As such it is a powerful tool to visually inspect the data. Tweets will be merged into clusters, depending on how far the user zooms into the map. At the most granular level, markers with individual Tweets are displayed that can be inspected by clicking on them.

\begin{example}
    R> cluster_tweets(mytweets)
\end{example}

Plotting methods are also provided for already fitted LDA models. For this purpose, \texttt{Twitmo} comes with built-in support for the popular \texttt{LDAvis} package. These plots are useful for interactively exploring and interpreting the topics of the fitted LDA models. \texttt{Twitmo} will export fitted LDA models to a \texttt{LDAvis} compatible JSON string and start the \texttt{LDAvis}-server, reducing the effort to one line of code. 

\begin{figure}[ht]
    \centering
    \includegraphics[width=1\textwidth]{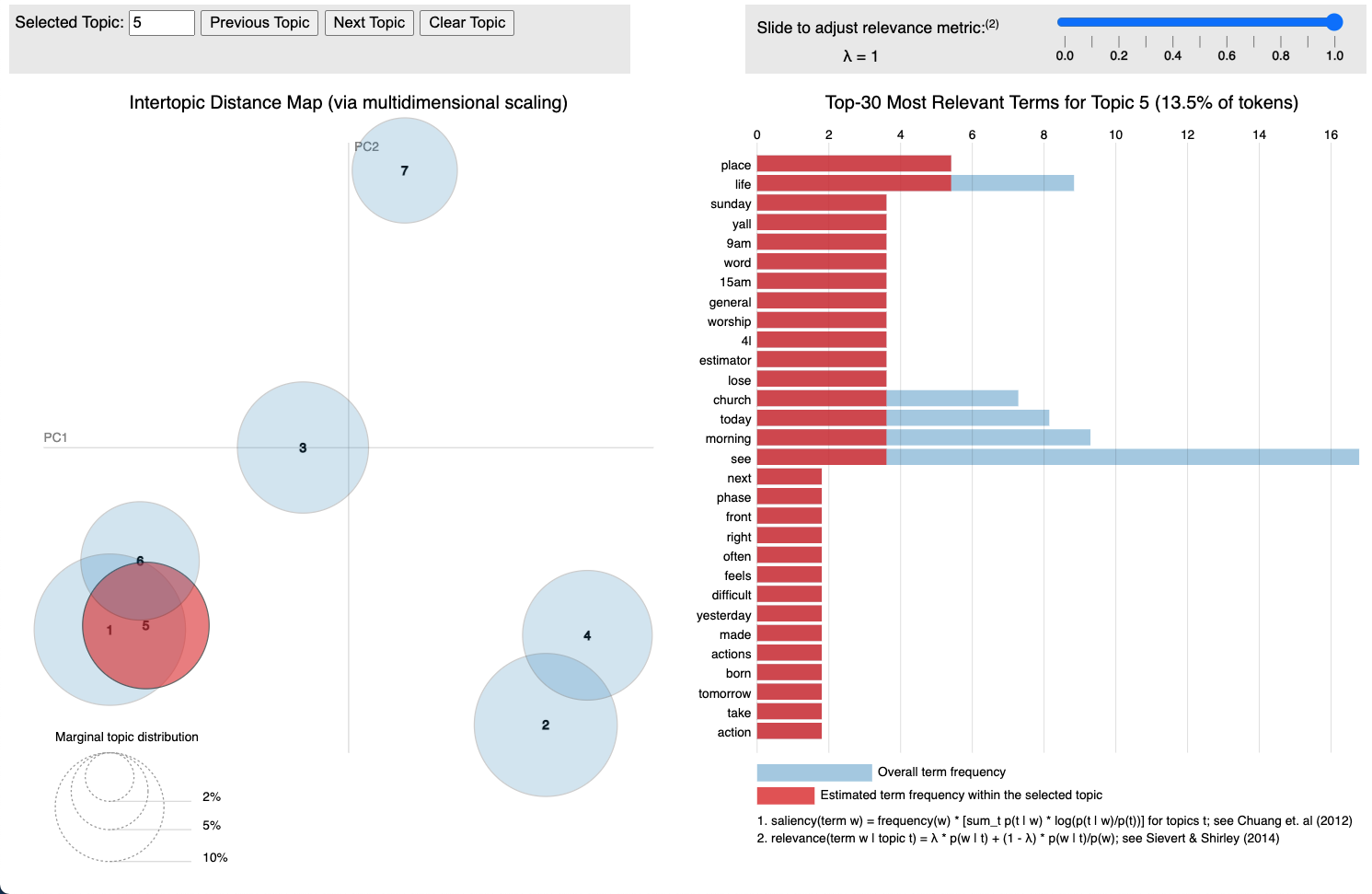}
    \caption{LDAvis plot of US mainland data fitted to 7-topic LDA model}
    \label{fig:ttvis_to_ldavis3}
\end{figure}

\begin{example}
    R> to_ldavis(model, pool$corpus, pool$document_term_matrix)
\end{example}

This can also be done for STMs via the \code{toLDAvis()} function provided by the \texttt{stm} package.

\begin{example}
    R> library(stm)
    R> toLDAvis(stm.model, docs = stm.model$prep$documents)
\end{example}

\section{Summary}

We introduced \texttt{Twitmo}, a topic modeling and visualization package tailored for working with geo-tagged Twitter data. We briefly reviewed currently available R packages for this purpose and show how \texttt{Twitmo} improves upon current packages in terms of usability and convenience, while also offering new functionality by providing a remedy to the sparsity and noisiness of Tweets. We also shortly review LDA and STM models and show how \texttt{Twitmo} can be used for fitting these models with Twitter data. Our package makes it easy to sample, pre-process and subsequently analyse the contents of geo-tagged Tweets and visualize the data as well as the results.
We improve the sampling process of geo-coded Tweets by supplying bounding box coordinates for 495 countries, regions and cities. The package can also detect latitude and longtide coordinates and convert them into a bounding box if given a radius.
For pre-processing we offer a wide range of methods, like stopword removal (including Twitter specific stopwords like "amp" or "rt"), symbol removal, removal of emojis, removal of punctuation, removal of URLs as well as removal of hashtags and usernames from the corpus and the option for creating n-grams.
With our package, Tweets can be pooled into longer pseudo-documents with our hashtag pooling function to produce coherent LDA topic models despite the shortness, sparsity and noisiness of Tweets. Tweet-level metadata can also be used to produce STM topic models as proposed by \cite{stm-long}.
Every piece of data parsed by our package can be visualized in static and interactive maps. For LDA models, our package offers conversion into \texttt{LDAvis} compatible JSON strings that is useful for model inspection.
By demonstrating the capabilities of \texttt{Twitmo} on a sample piece of data, which is included in \texttt{Twitmo}, we show how our package can be used to fit geo-tagged Twitter data into topic models and subsequently visualize the results of such an analysis.
\texttt{Twitmo} package is an innovative tool to work with geo-tagged Twitter data. The package may be extended to support a plethora of topic models in the future.

%Bibliography
\bibliographystyle{plainnat}  
\bibliography{twitmo}

\end{document}